\documentclass[12pt,a4paper]{article}

\usepackage{amssymb,amsmath}
\usepackage{color}
\usepackage{graphicx,geometry}

\newcommand{\R}{\mathbb{R}}

\newcommand{\minus}[1]{\mbox{$ - #1$}}

\newcommand{\nn}{\nonumber}

\begin{document}
{\center \textbf{{\Large On the incompleteness of the classification of quadratically integrable Hamiltonian systems in the three--dimensional Euclidean space}}\\[3mm]
{\large Libor {\v{S}}nobl$^1$}\\[3mm]
{\footnotesize $^1$ Czech Technical University in Prague, Faculty of Nuclear Sciences and Physical Engineering, \\
Department of Physics, B\v rehov\'a 7, 115 19 Prague 1, Czech Republic\\
Libor.Snobl@fjfi.cvut.cz\\}
}
{ \begin{center} {\large{\today}} \end{center} }

\begin{abstract}
We present an example of an integrable Hamiltonian system with scalar potential in the three--dimensional Euclidean space whose integrals of motion are quadratic polynomials in the momenta, yet its Hamilton--Jacobi / Schr\"odinger equation cannot separate in any orthogonal coordinate system. This demonstrates a loophole in the derivation of the list of quadratically integrable Hamiltonian systems in [Makarov et al., A systematic search for nonrelativistic systems with dynamical symmetries. Nuovo Cimento A Series 10, 52:1061--1084, 1967] where only separable systems were found, and the need for its revision.
\medskip

\noindent {\it Keywords\/}: {integrability; separability; natural Hamiltonians}
\end{abstract}

\section{Introduction}

In their seminal paper~\cite{MaSmoVaWin} Makarov, Smorodinsky, Valiev and Winternitz presented a list of quadratically integrable natural Hamiltonian systems in the three--dimensional Euclidean space and identified them with systems separable in ortogonal coordinate systems, cf.~\cite{EisenhartAnnMath,EisenhartPhysRev}. Their result is one of the standard references in the theory of integrable and superintegrable systems and lead to numerous further developments, see e.g. \cite{Evans1,Cleary} and the review~\cite{MiPoWin} for many others. It was widely accepted as a proof of the equivalence of quadratic integrability and separability in Euclidean 3D space, cf.~\cite{DaMajEng,BeWin} or the monograph~\cite{MillerBook1977}, p. 163. However, it became forgotten that the derivation of the list in~\cite{MaSmoVaWin} was based on one technical assumption which limits the universality of the above mentioned equivalence. In this short note we shall demonstrate that without that assumption a quadratically integrable however nonseparable 3D natural Hamiltonian system exists and the original derivation needs to be revisited.

\section{Review of the original argument and its loophole}

Let us review the argument used in~\cite{MaSmoVaWin} (with some streamlining due to later developments) and indicate the point where the analysis becomes incomplete. We consider the natural Hamiltonian for a particle of unit mass moving in the three--dimensional Euclidean space under the influence of the potential $V(\vec x)$,
\begin{equation}\label{Hamiltonian}
H= \frac{1}{2} \vec p + V(\vec x),
\end{equation}
and assume that it is integrable with a pair of integrals of motion $X_1$, $X_2$ which are quadratic polynomials in the momenta (henceforth abbreviated to \emph{quadratic integrals}), with coordinate dependent coefficients. For the sake of simplicity we shall proceed classically; however, the determining equations for the quadratic integrals and their solution are exactly the same in quantum mechanics, assuming total symmetrization of any terms involving noncommuting operators $\hat x_a,\hat p_a$.

To fix our notation, let us assume that the position vector in the Cartesian coordinates is expressed as $\vec x=(x,y,z)$, the canonically conjugated momenta to $x,y,z$ are denoted by  $p_x,p_y,p_z$ and the angular momenta are expressed as $l_x= y p_z- z p_y$, $l_y= z p_y-x  p_z$ and $l_z=x p_y-y p_x$.

As a consequence of the assumed form of the integrals, the Poisson brackets $\{H,X_1\}_{P.B}$, $\{H,X_2\}_{P.B}$ and $\{X_1,X_2\}_{P.B}$ are third order polynomials in the momenta $p_x,p_y,p_z$. As the Hamiltonian~\eqref{Hamiltonian} is an even polynomial in the momenta, the odd and even order terms in the integrals commute with the Hamiltonian~\eqref{Hamiltonian} independently and the integrals can be without any loss of generality assumed to be even or odd polynomials in the momenta. As any first order integral implies the existence of a second order integral as its square, we can assume that 
$X_1$ and $X_2$ are second order even polynomials in the momenta. The left hand sides of the involutivity conditions 
\begin{eqnarray}
\{H,X_1\}_{P.B} & = & 0, \quad \{H,X_2\}_{P.B}=0, \label{HX1HX2} \\
\label{involutivity}
\{X_1,X_2\}_{P.B} & = & 0
\end{eqnarray}
then become third order odd polynomials in the momenta. As the momenta are arbitrary, all their coefficients must vanish.

The third order terms in~(\ref{HX1HX2}--\ref{involutivity}) are easily solved and imply that the second order terms in $X_1$ and $X_2$ must be commuting elements in the universal enveloping algebra of the Euclidean algebra $\mathfrak{U}(\mathfrak{e}_3)$, i.e. quadratic polynomials in the linear and angular momenta. As we may arbitrarily combine the integrals with the Hamiltonian and among themselves, and the systems related by Euclidean transformations are physically equivalent, the leading order terms must belong to any of the classes of three--dimensional Abelian subalgebras consisting of quadratic elements in the universal enveloping algebra of the Euclidean algebra $\mathfrak{U}(\mathfrak{e}_3)$, which were recently classified in~\cite{MarSno22}.

It remains to solve the remaining conditions, which come from linear terms in the momenta in (\ref{HX1HX2}--\ref{involutivity}), namely to determine the scalar terms in the integrals $X_1$ and $X_2$, denoted by $m_1(\vec x)$ and $m_2(\vec x)$ below, and find the restrictions on the potential $V(\vec x)$ implied by their existence. The conditions coming from~\eqref{HX1HX2} are easily solved with respect to the first order derivatives of $m_1(\vec x)$ and $m_2(\vec x)$. Substituting these into~\eqref{involutivity}, one arrives at a set of three equations which are homogeneous linear first order PDEs for the potential $V(\vec x)$, cf.~\eqref{PBX1X2} below. As the coefficients of $\partial_a V$, $a=x,y,z$ form an antisymmetric $3 \times 3$ matrix $R$, it either has rank 2 or vanishes identically. At this point the authors of~\cite{MaSmoVaWin} stated
``Thus the potential V either satisfies three first-order equations -- a case which
will be considered separately -- or the consistency conditions (39) are satisfied
trivially.'' and proceeded assuming that the condition~\eqref{involutivity} vanishes identically. Only under this assumption they arrived at their list of quadratically integrable natural Hamiltonian systems and showed that one by one they precisely match with the separable systems of Eisenhart~\cite{EisenhartAnnMath,EisenhartPhysRev}.

The authors of~\cite{MaSmoVaWin} left several problems to be resolved in the planned Part II of their paper and we can assume that that's where they intended to address the case of the matrix $R$ of rank 2. However, due to external influence of political nature (military occupation of Czechoslovakia by the forces of Soviet Union and its satellites in 1968 and subsequent emigration of P. Winternitz to the other side of Iron Curtain) the authors' team split up and the sequel to~\cite{MaSmoVaWin} was never written.

The long forgotten assumption on the rank of the matrix $R$ came back to light recently, when we discussed with P. Winternitz the modification of the classification of quadratically integrable systems when linear terms in the momenta are present in the Hamiltonian, cf.~\cite{MSW,MS,MSW2}. P. Winternitz decided to revisit this question and assigned his student H. Abdul--Reda to work on it, resulting in the Master Thesis~\cite{HasanRedaThesis}. Its author concluded that the case of ${\mathrm{rank}} \, R=2$ does not lead to any new system and that it implies that the potential must be invariant under a two--dimensional Abelian subgroup of the Euclidean group, i.e. the system must possess two commuting first order integrals of motion. However, due to P. Winternitz's  demise in 2021 these results were never submitted for a journal publication and thus never passed validation by an independent peer review. Thus we decided to investigate the problem of ${\mathrm{rank}} \, R=2$ from the perspective of algebraic classification of leading order terms obtained with A. Marchesiello in~\cite{MarSno22} and, as we shall elucidate in the next section, arrived at the conclusion contradictory to that of H. Abdul--Reda, namely we find that quadratically integrable nonseparable system does exist.

\section{Quadratically integrable nonseparable system}
Let us look for quadratically integrable Hamiltonian systems~\eqref{Hamiltonian} of the form corresponding to the class (c) of~\cite{MarSno22}, namely with the integrals of motion of the form 
\begin{eqnarray}\label{IntAnsatz}
\nonumber  X_1 & = & l_x^2+l_y^2+l_z^2+ 2 b ( l_x p_x- (3 a-1) l_y p_y-2 l_z p_z)+\\
& & + 3  b^2 ( (1-4 a) p_x^2 - (3 a^2-2 a-1) p_y^2+ 2 (a-1) p_z^2)+m_1(\vec x),\\ 
\nonumber   X_2 & = & a l_y^2+l_z^2+6 a b l_x p_x+9 a b^2 (a p_z^2+p_y^2)+m_2(\vec x) , \quad 0< a \leq \frac{1}{2}, b\in\mathbb{R} \minus \{0 \}.
\end{eqnarray}
As the determinant $\det \frac{\partial (H, X_1,X_2)}{\partial( p_x,p_y,p_z)}$ does not identically vanish for any choice of parameters $a$ and $b$, the assumed form of the integrals~\eqref{IntAnsatz} implies functional independence of $H$, $X_1$ and $X_2$ (if they exist).

Let us recall from~\cite{MarSno22} that this form of the integrals represents one of the three possibilities for a pair of integrals such that both integrals involve terms quadratic in angular momenta. The allowed transformations, i.e. linear combinations of the integrals (and the Hamiltonian) and Euclidean transformations were all used in fixing the form of~\eqref{IntAnsatz}, namely
\begin{itemize}
\item the rotations were used to bring the terms quadratic in angular momenta in $X_2$ to a diagonal form and choose the range for the parameter $a$,
\item the translations were used to eliminate antisymmetric terms of the form $l_a p_b- l_b p_a$, $a,b=x,y,z$, from $X_1$,
\item the linear combinations of $X_1$ and $X_2$ were used to put the terms quadratic in angular momenta in $X_1$ to the simple form $\vec l^2$ and to eliminate $l_x^2$ term from $X_2$, the addition of the Hamiltonian was used to normalize the terms quadratic in the linear momenta in both $X_1$ and $X_2$,
\end{itemize}
(for a detailed derivation see pages 5--7 in~\cite{MarSno22}). Thus, we have no available transformations left and the assumed form of the integrals~\eqref{IntAnsatz} is not equivalent to any other known one, e.g. to the ones known from~\cite{MaSmoVaWin} (where the ones with both integrals quadratic in angular momenta correspond to spherical, oblate / prolate spheroidal, conical and ellipsoidal separation).

As a consequence of the assumed form of the integrals, i.e. the leading order terms of~\eqref{IntAnsatz} forming an Abelian subalgebra of $\mathfrak{U}(\mathfrak{e}_3)$, the Poisson brackets $\{H,X_1\}_{P.B}$, $\{H,X_2\}_{P.B}$ and $\{X_1,X_2\}_{P.B}$ reduce to first order polynomials in the momenta $p_x,p_y,p_z$, without zeroth order terms. Separating the conditions~\eqref{HX1HX2} into coefficients of $p_x,p_y,p_z$ and solving them with respect to the first order derivatives of $m_1(\vec x)$ and $m_2(\vec x)$ we find:
\begin{eqnarray}\label{m1m2}
\nn \partial_x m_1(\vec x) & = & 2 ( 3(1-4 a ) b^2 + y^2 + z^2) \partial_x V(\vec x) - 2 (3 a b z + x y) \partial_y V(\vec x) \\ & & \nn + 2 (3 b y - x z) \partial_z V(\vec x), \\
\nn  \partial_y m_1(\vec x) & = & -2 (3 a b z + x y) \partial_x V(\vec x) + 2 ( 3 (1 + 2 a -3 a^2) b^2 + x^2 + z^2) \partial_y V(\vec x) \\ & & \nn  - 2 ( 3 b ( 1-a )x  +y z) \partial_z V(\vec x),\\
\nn  \partial_z m_1(\vec x) & = & 2 (3 b y - x z) \partial_x V(\vec x) - 2 (  3 b (1-a) x+ y z) \partial_y V(\vec x) \\ & & \nn  + 2 (6 (a-1) b^2 + x^2 + y^2) \partial_z V(\vec x),\\
  \partial_x m_2(\vec x) & = & 2 (a z^2 + y^2) \partial_x V(\vec x) - 2 (3 a b z + x y) \partial_y V(\vec x) \\ & & \nn  + 2 a (3 b y - x z) \partial_z V(\vec x), \\
\nn  \partial_y m_2(\vec x) & = & -2 (3 a b z + x y) \partial_x V(\vec x) + 2 (9 a b^2 + x^2) \partial_y V(\vec x), \\
\nn  \partial_z m_2(\vec x) & = & 2 a (3 b y - x z) \partial_x V(\vec x) + 2 a (9 a b^2 + x^2) \partial_z V(\vec x).
\end{eqnarray}
Their compatibility implies a set of second order linear PDEs for the potential $V(\vec x)$. On the other hand, substituting~\eqref{m1m2} into~\eqref{involutivity} we obtain a set of three first order linear homogeneous PDEs for the potential $V(\vec x)$, out of which two are independent and can be written as
\begin{eqnarray}\label{PBX1X2}
\nn \frac{\partial_x V(\vec x)}{\partial_z V(\vec x)} & = & \frac{(1-a)\left(9 a(a-1) b^2 x - 6 a b y z + a x(x^2 +  z^2) - x (x^2 + y^2)\right)}{a \left(9 (a-1) a b^2 z + 6 a b x y - a z(x^2 +z^2) - 6 b x y + z(x^2  - y^2)\right)},\\
\frac{\partial_y V(\vec x)}{\partial_z V(\vec x)} & = & -\frac{9 a(a-1) b^2 y + 6 a (1-a) b x z  + a y (z^2 - x^2 ) + y (x^2 + y^2)}{a \left(9 a (a-1) b^2 z + 6 (a-1)  b x y - a z (x^2 +  z^2)  + z (x^2 - y^2 )\right)}.
\end{eqnarray} 
Notice that the numerators and denominators on the right hand side of~\eqref{PBX1X2} cannot identically vanish for any choice of the parameters $a$ and $b$ due to $x y^2$, $x y^2$ and $y^3$ terms. Thus we are indeed considering the case with $\mathrm{rank}\, R=2$ for any choice of the parameters.
Solving~\eqref{PBX1X2} using the method of characteristics we find that $V(\vec x)$ must be an arbitrary function $v(u)$ of an invariant coordinate, which can be conveniently chosen as
\begin{eqnarray}\label{PBX1X2sol}
 u & = & (a - 1)^2 x^4 + (a z^2 + y^2)^2 + 2 (1 - a) x^2 (y^2-a z^2)  + \\
\nn & & 6 a b (a - 1) \left(3 \left((x^2 - z^2) a - x^2 + y^2\right) b - 4 x y z\right)+81 a^2 (1 - a)^2 b^4.
\end{eqnarray}
Substituting $V(\vec x)=v(u)$ into the compatibility conditions for~\eqref{m1m2} we find a system of ODEs which reduces to a single equation
\begin{equation}
2 u \frac{{\mathrm d}^2 v(u)}{ {\mathrm d} u^2} = -3 \frac{{\mathrm d} v(u)}{ {\mathrm d} u}.
\end{equation}
Ignoring the irrelevant additive constant in the potential, we find the potential determined up to a multiplicative constant $w_0$,
\begin{equation}\label{Vsolved}
V(\vec x)=v(u)=\frac{w_0}{\sqrt{u}}.
\end{equation}
In the next step, the equations~\eqref{m1m2} determine the scalar terms $m_1$ and $m_2$ in the integrals up to irrelevant additive constants. They read
\begin{eqnarray}\label{m1m2solved}
\nn m_1(\vec x) & = & 2 w_0 \frac{ x^2 +y^2+z^2 + 3 b^2 (1-a)}{\sqrt{u}}, \\
m_2(\vec x) & = & w_0 \frac{x^2 + y^2 + a (x^2 + z^2)+9 a b^2 (a+1)}{\sqrt{u}},
\end{eqnarray}
where $u$ is the quartic polynomial in the coordinates introduced in~\eqref{PBX1X2sol}.
\medskip

The system with the potential~\eqref{Vsolved} is an integrable system with the integrals of the form~\eqref{IntAnsatz} where the functions $m_1$ and $m_2$ are specified by~\eqref{m1m2solved}. An explicit but somewhat lengthy computation shows that it does not possess any other integrals of motion linear or quadratic in the momenta. Thus it cannot intersect any other class of integrable systems of~\cite{MarSno22} and conseqently it cannot be transformed using Euclidean transformations to any of the quadratically integrable and separable systems of~\cite{MaSmoVaWin}. Therefore, it provides an example of a quadratically integrable yet not separable natural Hamiltonian system.

Let us mention that the polynomial~\eqref{PBX1X2sol} can of course  have real roots and thus the potential~\eqref{Vsolved} may blow up in the configuration space. Looking up its roots we find that the polynomial~\eqref{PBX1X2sol} vanishes along the complex hyperplanes 
\begin{equation}\label{singular_hyperplanes}
\epsilon_1 \sqrt{1 - a} \mathrm{i} x + \epsilon_1 \epsilon_2 y + \epsilon_2 \sqrt{a}  \mathrm{i} z = 3 \sqrt{a (1 - a)} b, \qquad \epsilon_1,\epsilon_2=\pm 1,
\end{equation}
i.e. the real polynomial~\eqref{PBX1X2sol} factorizes over the complex field as
\begin{equation}
u=\prod_{\epsilon_1=\pm 1,\epsilon_2=\pm 1  } \left( \epsilon_1 \sqrt{1 - a} \mathrm{i} x + \epsilon_1 \epsilon_2 y + \epsilon_2 \sqrt{a}  \mathrm{i} z - 3 \sqrt{a (1 - a)} b\right)
\end{equation}
and the potential~\eqref{Vsolved} can be expressed as
\begin{equation}\label{VsolvedA}
V(\vec x)=\frac{w_0}{\sqrt{\prod_{\epsilon_1=\pm 1,\epsilon_2=\pm 1  } \left( \epsilon_1 \sqrt{1 - a} \mathrm{i} x + \epsilon_1 \epsilon_2 y + \epsilon_2 \sqrt{a}  \mathrm{i} z - 3 \sqrt{a (1 - a)} b\right)}}.
\end{equation}
As both the parameters $a,b$ and the coordinates $x,y,z$ are real (and $0<a\leq \frac{1}{2}$), restricting the hyperplanes~\eqref{singular_hyperplanes} to the real domain we see that $V(\vec x)$ blows up along the two straight lines given by 
\begin{equation}\label{singularities}
 x =- \epsilon_1 \epsilon_2 \sqrt{ \frac{a}{1 - a}} z  , \quad  y = 3 \epsilon_1 \epsilon_2 \sqrt{a (1 - a)} b, \qquad \epsilon_1,\epsilon_2=\pm 1.
\end{equation}
As these do not separate $\R^3$ into disconnected domains, everywhere else the potential~\eqref{VsolvedA} is a well--defined real function.

If we assume that the parameter $w_0$ is positive, the singular lines~\eqref{singularities} are not dynamically accessible for any initial condition with finite energy. Thus our Hamiltonian system is a well--defined integrable system in the configuration space defined as $\R^3$ without the two singular lines~\eqref{singularities}. Whether the singularities are dynamically reachable for negative values of $w_0$ in finite time we don't know yet.

\begin{figure}
\resizebox{0.7\hsize}{!}{\includegraphics{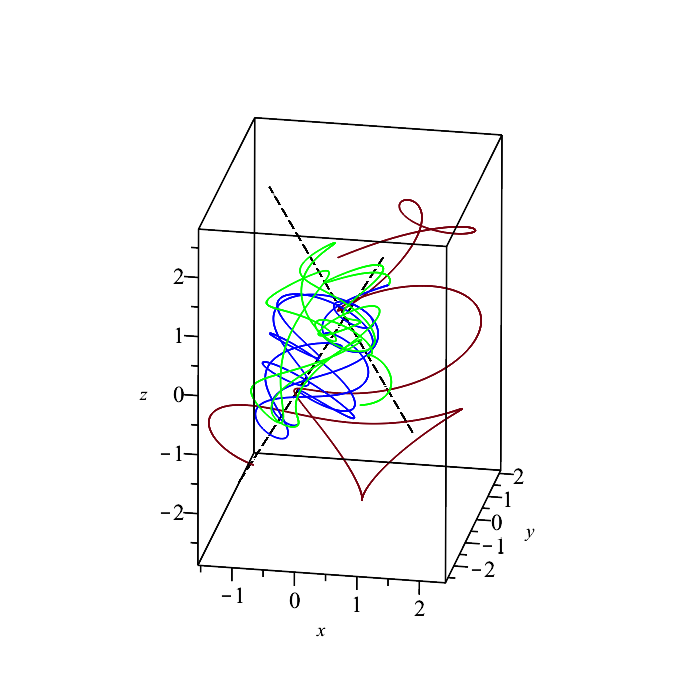}}\\\resizebox{0.7\hsize}{!}{\includegraphics{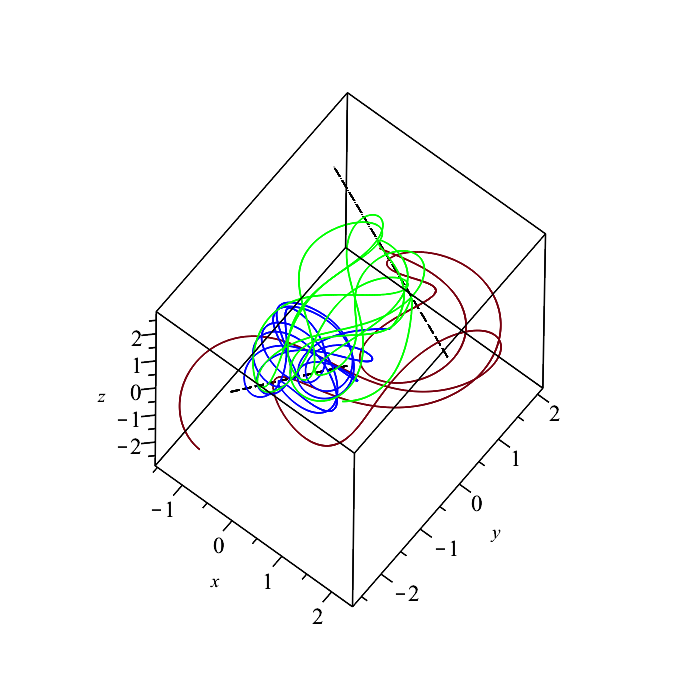}}
\caption{Three examples of bounded (i.e. negative energy) trajectories with the parameters $a=\frac{1}{4}$, $b=1$, $w_0 = -1$ viewed from two different directions.  The lines along which the potential becomes infinite are also indicated, dashed.}
\end{figure}

\section{Conclusions}
The purpose of this short note is to bring the research community's attention to the forgotten assumption in~\cite{MaSmoVaWin} and explicitly demonstrate that the statement on the equivalence of quadratic integrability and separability in 3D Euclidean space does not hold if that assumption is violated, arriving at a new quadratically integrable yet not separable system with the potential~\eqref{VsolvedA}. Let us recall that as the determining equations for the quadratic integrals of motion of natural Hamiltonian systems~\eqref{Hamiltonian} in classical and quantum mechanics coincide, also the quantum system with the potential~\eqref{VsolvedA} is integrable with integrals of motion quadratic in momenta.

 As we have already mentioned in the Introduction, P. Winternitz, one of the original authors of~\cite{MaSmoVaWin}, learned about the possible loophole in~\cite{MaSmoVaWin} and was aiming to fill it in with his student in the thesis~\cite{HasanRedaThesis}. Our system should have been found there in Section 2.3.1, as it is of the form closely resembling (2.3.5)  therein (up to typos), but it wasn't. As we have seen above, its existence invalidates the two conclusions of~\cite{HasanRedaThesis}, namely that the case of ${\mathrm{rank}}\, R=2$ does not lead to any new integrable systems and that it implies that any integrable  system with ${\mathrm{rank}}\, R=2$ must possess two commuting first order integrals of motion.

While discussing a preliminary version of this note with G. Rastelli, he pointed out that our result can also be viewed from a geometrical perspective. Namely, the leading order terms of quadratic integrals can be encoded into Killing 2--tensors and by theorem of Eisenhart in~\cite{EisenhartAnnMath} (see also \cite{KalMil,BCR2000} or the monograph~\cite{KalKreMilBook}, Sec. 4.3.1, for its recent formulations and discussion) the construction of separable orthogonal coordinates requires not only the integrals of motion to be in involution but also the existence of simultaneous eigenvectors of both Killing tensors (after lowering one of their indices by the metric which allows to interpret them as linear operators on 1--forms). As G. Rastelli observed the assumed form of the integrals~\eqref{IntAnsatz} implies Killing tensors that do not commute, i.e. do not possess simultaneous eigenforms. He also pointed out that as the Killing tensor associated to the integral $X_1$ does not possess any Killing vectors, nonorthogonal separation of coordinates is also not possible. From this point of view our result means that a nonconstant potential in the 3D Euclidean space exists such that it possesses a pair of quadratic integrals in involution whose Killing tensors do not commute and no other pair of quadratic integrals for it exists. 

It is not yet known whether our system~\eqref{VsolvedA} is the sole exception to the statements in~\cite{MaSmoVaWin,HasanRedaThesis} or whether other quadratically integrable nonseparable systems in Euclidean 3D space do exist. Thus, a complete re--derivation of the list of quadratically integrable systems based on the classification of the leading order terms in~\cite{MarSno22} is currently under way and we expect to report on it in not too distant future. We also intend to study the presented system~\eqref{VsolvedA} in more detail, e.g. attempt to find its action--angle variables, as well as to analyse its quantum counterpart.

\section*{Acknowledgements}
The author received support from the Czech Ministry of Education, Youth and Sports under RVO 68407700. He thanks Antonella Marchesiello and Giovanni Rastelli for discussions on the first versions of the manuscript.


\end{document}